\documentclass[twocolumn,english,aps,prb,twocolum,superscriptaddress,bibnotes,amsmath,amssymb,floatfix]{revtex4-2}

\usepackage[colorlinks=true,citecolor=blue,linkcolor=magenta]{hyperref}
\usepackage{changes}
\usepackage{lineno}
\usepackage{stmaryrd}
\usepackage{bm}
\usepackage{graphicx}
\usepackage{epsfig}
\usepackage{color}
\usepackage{amssymb}
\usepackage{amsmath}
\usepackage{array}
\usepackage{braket}
\usepackage{soul}
\usepackage{cleveref}
\usepackage[caption=false]{subfig}
\usepackage[english]{babel}
\usepackage{letltxmacro}
\usepackage{siunitx}
\usepackage{dsfont}
\usepackage{verbatim}
\usepackage{gensymb} 

\def\ii{{\rm i}}  \def\ee{{\rm e}}
\def\rb{{\bf r}}  \def\Rb{{\bf R}}    \def\vb{{\bf v}}
    \def\zz{\hat{\bf z}}    
\def\kb{{\bf k}}      
      
\def\me{m_{\rm e}}  
\def\Eb{{\bf E}}    \def\Ab{{\bf A}}

   \def\PM{{P\!M}}

\newcommand{\MPINAT}{Max Planck Institute for Multidisciplinary Sciences, D-37077 G\"{o}ttingen, Germany}
\newcommand{\UGOE}{4th Physical Institute - Solids and Nanostructures, University of Göttingen, D-37077 G\"{o}ttingen, Germany}

\begin{document}


\title{
Femtosecond and attosecond phase-space correlations in few-particle photoelectron pulses
}

\author{Rudolf Haindl}
\affiliation{\MPINAT}
\affiliation{\UGOE}
\author{Valerio Di Giulio}
\email{valerio.digiulio@mpinat.mpg.de}
\affiliation{\MPINAT}
\affiliation{\UGOE}
\author{Armin Feist}
\affiliation{\MPINAT}
\affiliation{\UGOE}
\author{Claus Ropers}
\email{claus.ropers@mpinat.mpg.de}
\affiliation{\MPINAT}
\affiliation{\UGOE}


\date{\today}

\begin{abstract}
Temporal correlations in pulsed electron beams reflect the microscopic dynamics of emission and interparticle interaction. In femtosecond electron emission from nanoscale field emitters, Coulomb interactions result in structured few-electron states with strong correlations in energy, time, and transverse momentum. Interactions with external fields may be used to both probe and further manipulate these correlated states.
Here, we combine femtosecond-gated, event-based detection with inelastic electron-light scattering to directly map the photoelectron phase-space distribution of two-electron states. Our experiments demonstrate a bimodal structure in longitudinal phase space, with distinct contributions from interparticle interaction and dispersion. Moreover, we theoretically reveal that global phase modulation coherently shapes the few-electron phase-space distribution to exhibit attosecond temporal correlations. This controlled phasing of few-electron states can be harnessed to produce tailored excitations and super-radiance via two-electron energy post-selection.

\end{abstract}

\maketitle

\clearpage

The Coulomb interaction among electrons forming a beam remains a longstanding challenge for electron microscopy and spectroscopy. Limiting contrast and resolution, longitudinal~\cite{Boersch1954} and transverse~\cite{Loeffler1969a} scattering manifests as anticorrelations in the time of arrival~\cite{Borrelli2024} as well as the transverse position and momentum~\cite{Kiesel2002, Kodama2011, Kuwahara2021, Keramati2021, Haindl2023a} at the detector plane. In pulsed electron beams produced by photoemission from nanotip field-emitters~\cite{Hommelhoff2006d, Ropers2007a, Barwick2007}, event-based detection recently uncovered pronounced Coulomb-induced energy correlations in an electron microscope~\cite{Haindl2023a} and at free-standing nanotips~\cite{Meier2023}. The understanding of such correlations promote novel measurement schemes based on controlled few-electron states, underlying recent proposals of shot noise-reduced imaging~\cite{Haindl2023a, Koppell2025} and multi-particle sample excitations~\cite{Gover2020, Zhao2021a, Karnieli2021, GarciadeAbajo2021, Ratzel2021, Zhang2021, Ruimy2021, Kfir2021a, DiGiulio2021, Gorlach2024, DHR25}, where excitation enhancement is achieved either via pulse compression or periodic bunching. Moreover, multi-electron states are expected to enhance dose-sensitive imaging~\cite{VandenBussche2019, Axelrod2024} and quantum-mechanical electron-light entanglement~\cite{Kfir2019, Talebi2021, Konecna2022a, Kazakevich2024a, Henke2025a, Henke2025b,Preimesberger2025}.

The phase-space distribution of pulsed electron beams is particularly relevant in the context of ultrafast electron microscopy, combining high temporal and spatial resolutions~\cite{Zewail2010, Feist2017, Bach2019, Olshin2020, Zhu2020a, Shiloh2022, Kuttruff2024a, Schroder2024}. On the one hand, typically low duty cycles in ultrafast imaging, diffraction and spectroscopy aim for maximum allowable pulse charge, which renders Coulomb interactions an essential constraint in this technology. On the other hand, ultrafast microscopes offer ideal conditions to study the correlations among few-electron ensembles within a well-defined temporal gate.  Several effects govern the spectral and temporal distribution of electron pulses, which involve the photoemission process~\cite{Hommelhoff2006d, Ropers2007a, Barwick2007, Dowell2009, Cook2009, Herink2012, Kruger2011, Hergert2024} as well as Coulomb interactions~\cite{Cook2016, Kuwahara2016, Bach2019, Meier2022a, Haindl2023a}. Measurements of the time structure of electron pulses can be obtained by electron-laser cross-correlation using ponderomotive scattering~\cite{Hebeisen2008} or streaking~\cite{Kozak2017, Epp2024}, recently performed for few-electron states \cite{Kuttruff2024a}. Femtosecond-resolved temporal slicing~\cite{Park2012, Kirchner2014, Feist2017, Bach2019} based on inelastic electron-light interaction~\cite{Barwick2009, GarciadeAbajo2010, Feist2015} has shown a broadening of the phase-space distribution with increasing average current. However, mean-field and stochastic interactions are expected to affect pulse dispersion in distinct ways, which calls for the measurement and active shaping of electron-number specific phase-space distributions for tailored probing.

\begin{figure}[!b]
\centering
\includegraphics[width=\columnwidth]{figures/phase_space_manuscript/fig_1/fig1_new_v34.pdf}
\caption{Generation and propagation of two-electron pulses in a field-emitter Ultrafast Transmission Electron Microscope (UTEM). (a) Schematic of the increasing temporal separation of a Coulomb-correlated photoelectron pair (left) accelerated to \SI{200}{keV} energy (right) and recorded using an event-based detector behind an electron spectrometer. Note the logarithmic position scale during acceleration. (b) Simulated evolution of the longitudinal phase-space distribution for different locations along the microscope column (indicated by i-iv), illustrating the concurrent buildup of energy correlations and dispersive pulse broadening.}
\label{fig_1}
\end{figure}

In this Letter, we present the measurement and control of two-electron states using inelastic electron-light scattering (IELS). Employing femtosecond temporal gating in an ultrafast transmission electron microscope (UTEM), we record time-energy correlation maps of Coulomb-correlated two-electron pulses and reconstruct their joint phase-space distributions. To support these observations, we formulate IELS in the multi-electron regime and perform numerical simulations that reproduce the measured energy–time correlations at the sample plane. Finally, we demonstrate that coherent IELS enables the preparation of attosecond-scale interparticle correlations, allowing for the active control and enhancement of nanoscale excitations, even without relying on a dispersive bunching of the electron ensemble.

Optimized electron pulse properties are of central importance in ultrafast electron microscopy. In principle, the longitudinal properties of multi-electron ensembles are fully described by the 2$N$-dimensional phase-space density $\rho_N(\{t_i\},\{E_i\})$, where $\{t_i\}$ and $\{E_i\}$ denote the set of $N$ time and energy variables, respectively. Due to the high dimensionality of $\rho_N(\{t_i\},\{E_i\})$, marginal phase-space densities are often sufficient in the analysis of beam properties and the determination of achievable energy and time resolution. These may be expressed in terms of the overall densities $\rho(t, E)$, or number-sorted as $\rho_N(t, E)$. In particular, the former results from the marginalization of $\rho_N(\{t_i\},\{E_i\})$ with respect to the total number of electrons employed in the experiment, while the latter refers only to pulses containing $N$ electrons. 

This study aims at first characterizing and then optically shaping the number-sorted phase-space properties of few-electron pulses within a field-emitter UTEM (see Fig.~\ref{fig_1}a). In order to illustrate the Coulomb- and propagation-induced emergence of energetic and temporal correlations, we consider the results of two-electron trajectory simulations predicting the evolution of $\rho_2(t, E)$ along different positions in the microscope (see Appendix \ref{aA}). Upon photoemission, the initial conditions of the double-electron pulse in time and energy are normally distributed (Fig.~\ref{fig_1}b, panel i), that is, we assume no nascent correlations. Subsequently, the electrons are electrostatically accelerated along the beam direction. To a good approximation, the momentary rate of Coulomb-induced energy exchange is given by the Coulomb force times the mean velocity of the two electrons, which leads to an increasing difference in kinetic energy (panel ii). The external accelerating field decreases during propagation, the kinetic energy difference saturates, and dispersion translates the built-up energy correlation into a temporal correlation (panels iii \& iv).

\begin{figure}[!t]
\centering
\includegraphics[width=\columnwidth]{figures/phase_space_manuscript/fig_2/fig2_new_v7.pdf}
\caption{Inelastic electron-light scattering of few-electron pulses. a) Coulomb-correlated electron pulses traverse a time-delayed optical near field at a silicon membrane and undergo inelastic electron-light scattering. b-d) Pump-probe delay scans of pulse charge-resolved components $n=1$ (panel b), $n=2$ (panel c) and $n=3$ (panel d) for inelastic few-electron-light scattering. e-g) Plot of spectral density change obtained by subtracting the spectrum at large time delays. As the optical near field scatters the temporally overlapped electrons, this spectrogram unveils the temporal separation of electrons within a pulse. }
\label{fig_2}
\end{figure}

\begin{figure*}[!t]
\centering
\includegraphics[width=\textwidth]{figures/phase_space_manuscript/fig_3/fig_3_new_v27.pdf}
\caption{Phase-space reconstruction of double-electron states via inelastic electron-light scattering. a) Spectrum (top), scheme (middle) and electron energy pair histogram (bottom) of the bare Coulomb correlated $n=2$-state. b,c) Top: Upon interaction with a time-delayed short pump pulse, the faster electron (b) or the slower electron (c) of the $n=2$-state are selectively scattered at the optical near field, while the slower or faster electron is not scattered. Bottom: Inelastic electron-light scattering  pair histograms of energies $E_\mathrm{1/2}$ for coincident electrons A (the slower electron) and B (the faster electron) at pump-probe time delays of \SI{-100}{fs} (b) and \SI{100}{fs} (c) and simulated correlation histogram at both time delays. The green and pink rectangles represent the first and second electron's energy gain regions, respectively. f) Gain-filtered IELS spectra of faster (green) and slower (pink) electron of two-electron states.}
\label{fig_3}
\end{figure*}

In the transmission electron microscope (JEM F200, JEOL) (see Fig.~\ref{fig_1}a), we generate few-particle electron pulses by close-to-threshold photoemission from a Schottky emitter~\cite{Cook2009, Feist2017, Bach2019, Haindl2023a} with a femtosecond laser (\SI{2}{MHz} repetition rate, \SI{160}{fs} pulse duration and \SI{515}{nm} wavelength). The number of electrons emitted by each laser pulse follows moderately antibunched statistics~\cite{Haindl2023a}, with a power-dependent rate of multi-electron emission events and a transmission into the column of below 5\%. The electron pulses are accelerated to \SI{7}{keV} in an emitter assembly consisting of an extraction and focusing electrode and subsequently to a beam energy of $E_0=\SI{200}{keV}$ (corresponding to a speed $v\approx\SI{0.7}{c}$). For spectral analysis, we use an energy filter and an event-based camera with nanosecond time resolution based on the Timepix3 ASIC~\cite{Ballabriga2018}. Our detection scheme (see Ref.~\cite{Haindl2023a, vanSchayck2020} allows 
to correlate the incident electrons to the femtosecond laser pulses that generated them and thereby classify the electron pulses according to their electron number. 

The ensemble phase-space densities are reconstructed by ultrafast gating of the electron pulses. Conceptually, we relate $\rho_N(\{t_i\},\{E_i\})$ to measurable electron energy spectra by exploiting inelastic electron-light scattering (IELS), a process where electrons are efficiently coupled to optical near fields~\cite{Park2012, Feist2015, Feist2017, Bach2019}. Here, few-electron states arriving in coincidence with an out-of-plane-polarized optical near field in the sample plane at time $t_\mathrm{L}$ are inelastically scattered by multiples of the photon energy. In this regard, as shown in the Appendix \ref{aB}, the recorded spectral maps $\langle \Gamma(t_{\rm L},\{E_i\})\rangle$ can be written as the multi-dimensional convolution ($\cdot$ symbol)

\begin{align}
 \langle \Gamma(t_{\rm L},\{E_i\})\rangle=\frac{1}{N}P(t_{\rm L},\{E_{i}\}) \cdot \rho_N(t_{\rm L},\{E_{i}\}),  \label{conv1}
\end{align}
between the $N$-body phase-space density $\rho_N(\{t_{0,i}\},\{E_{0,i}\})$ and the IELS slicing function $P(\{t_{0,i}\},\{E_{0,i}\})=\prod_{i=1}^N \Gamma(t_{\rm L}-t_{0,i},E_i-E_{0,i})$ over the delay times $\{t_{0,i}\}$ and energies $\{E_{0,i}\}$ of the electrons entering the IELS scattering region.

We begin characterising the few-electron states by acquiring pump-probe spectrograms as sketched in Fig.~\ref{fig_2}a. In these experiments, we focus the nanometer-sized electron beam onto a silicon membrane with a thickness of \SI{35}{nm} and electron-laser and electron-sample angles $\theta_\mathrm{el./las.}=55\degree$ and $\theta_\mathrm{el./sam.}=10\degree$. An optical near-field is created under illumination by femtosecond laser pulses with peak intensities of \SI{11}{GW/cm^2}, a center photon energy $\hbar \omega_{\rm L} = \SI{1.72}{eV}$ and a full-width-at-half-maximum intensity duration of $\approx 1.67\Delta_\mathrm{L} \approx \SI{120}{fs}$. The electrons traversing the optical near-field at the membrane experience IELS with an electron-light coupling strength $|\beta|= 5.8$. Figure~\ref{fig_2}b-d shows optical-pump-electron-probe delay scans, sorted with respect to the number of electrons in the pulse. We observe an $n$-electron rate per laser pulse of $r_1=\SI{7e-2}{}$, $r_2=\SI{2.3e-3}{}$ and $r_3=\SI{4.6e-5}{}$. At large delays, we identify the characteristic single-, double- and triple-lobed spectra of $n=1-3$ electron states~\cite{Haindl2023a}. As the temporal overlap is approached, electrons are inelastically scattered to higher and lower energies, and the zero-loss peak is depleted. This is evident in the plot of the IELS-induced change in spectral density in Fig.~\ref{fig_2}e-g, defined as the delay-dependent spectrum subtracted by the spectrum at large delays $\langle \Gamma(t_{\rm L},\{E_i\})\rangle-\langle \Gamma(t^{\rm max}_{\rm L},\{E_i\})\rangle$. We identify a single--, double-- and triple-peaked region for number states $n=1-3$, showing that the electrons within a state have different arrival times at the sample plane, as observed in streaking experiments~\cite{Kuttruff2024a}. 

For a quantitative analysis, we consider the electrons that were scattered to high sideband orders near the peak intensity of the laser pulse. Generally, this approach can be adopted to any number of electrons with a well-defined energy separation. In particular, we investigate electron pulses containing two electrons. These states exhibit a spectral double-lobe feature with an energy gap of \SI{1.8}{eV} between the electrons, visible in the pair correlation histogram (see Fig.~\ref{fig_3}a). The general use of only a single laser pulse in IELS experiments, providing one single reference time $t_{\rm L}$, prevents the full reconstruction of the multi-particle correlations through $\rho_N(\{t_i\},\{E_i\})$ [see Eq. (\ref{conv1})]. However, its marginal $\rho_N(t,E)$ is accessible, as described in the following. 

The analysis simplifies because, experimentally, at most one of the two electrons undergoes IELS at any given delay. This follows from a sufficient temporal separation of the energetically separated electrons. We assign the initially slower electron, which arrives later at the sample, the energy $E_\mathrm{A}$, while the initially higher energy is denoted $E_\mathrm{B}$. An optical near field is generated at pump-probe delay $t_{\rm L}$ to selectively scatter either electron, as exemplified for negative and positive pump-probe delays of $t_{\rm L}=\SI{-100}{fs}$ (Fig.~\ref{fig_3}b, upper panel) and $t_{\rm L}=\SI{+100}{fs}$ (Fig.~\ref{fig_3}c, upper panel), respectively. Two-dimensional pair density maps of the two measured energies $E_1$ and $E_2$ after the interaction illustrate the selective interaction with the initially faster or slower electron, as shown in the lower panels of Fig.~\ref{fig_3}b\&c. In both cases, electrons are scattered from their initially separated energies (Fig.~\ref{fig_3}a) along the $E_\mathrm{1}$ and $E_\mathrm{2}$ axes. Importantly, the generated IELS combs exhibit delay-dependent offsets. Specifically, the green and pink open-ended rectangles act as selective electron-gain filters for the faster and slower electron, respectively. In our analysis, we identify gain-scattered contributions from these regions, with the extracted energy- and delay-dependent densities shown in Fig.~\ref{fig_3}f. The gain contributions are temporally separated, but overlap to some extent at small delays. Assuming laser-pulse durations short compared to the electron pulse profile, this procedure links the different contributions forming the $2$-dimensional phase-space density $\rho_2(t,E)=\rho_A(t,E)+\rho_B(t,E)$ to the relative intensities $I_A$ (pink) and $I_B$ (green) of the filtered spectrograms, now expressed by the convolutions
\begin{align}
I_{A/B}(t_{\rm L},E_{2})\approx \frac{1}{2}\Gamma(t_{\rm L},E_{2}) \cdot \rho_{A/B}(t_{\rm L},E_{2}),
\label{conv2}
\end{align}
(see Appendix \ref{aC} and Refs. \cite{Park2012, Plemmons2014, Feist2017, Bach2019}).
Therefore, the data analysis required to obtain $\rho$ simplifies to a deconvolution of Eq.~(\ref{conv2}), carried out by a standard procedure. We Fourier transform the gain-filtered electron density at each time delay along the spectral axis. Then, we extract the phase of the Fourier transformation at the Fourier frequency of the photon energy to obtain the spectral phase of the IELS modulation, representing the momentary chirp $E_{A/B}^\mathrm{chirp}(t)$. Together with the spectral width from the Gaussian envelope of the Fourier transform, this yields a slice of the two-electron phase-space distribution $\rho_2(t,E)$ at the electrons' arrival time in the sample plane (see Appendix \ref{aC}).

\begin{figure}[!t]
\centering
\includegraphics[width=\columnwidth]{figures/phase_space_manuscript/fig_4/fig_4_new_v18.pdf}
\caption{Reconstructed marginalized phase-space densities of single- (a\&c) and double-electron states (b\&d) at the sample plane for electron rates $r_1 =\SI{1.5e-2}{}, r_2=\SI{1.1e-4}{}$ (a\&b) and $r_1 =\SI{3.5e-2}{}, r_2=\SI{5.6e-4}{}$ (c\&d), respectively. The dashed lines are linear fits, revealing a single-electron pulse chirp of \SI{390}{fs/eV} and \SI{250}{fs/eV} for the two laser powers, respectively. The linear fits in b\&d yield reduced chirps of \SI{420}{fs/eV} (b) and \SI{300}{fs/eV} (d) of the sub-ensembles. In comparison, the slope between the two maxima of the phase-space distributions (solid lines) are \SI{130}{fs/eV} (b) and \SI{140}{fs/eV} (d), respectively.}
\label{fig_4}
\end{figure}
The resulting phase-space densities of the single and double-electron pulses shown in Fig.~\ref{fig_4} are clearly distinct. While the single-electron distribution exhibits one lobe, the two-electron pulse distribution is bimodal with two regions spectrally and temporally separated by \SI{1.8}{eV} and \SI{240}{fs}, respectively (Fig.~\ref{fig_4}a\&b). In general, the phase-space densities are affected by mean-field and stochastic interaction close to the emitter~\cite{Feist2017, Bach2019}, resulting in an electron pulse chirp that depends on the photoemission current and is also affected by scattering with electrons not entering the beam (compare the upper and lower panels in Fig.~\ref{fig_4}). We find a similar chirp and current-dependent change thereof for the phase-space distribution of single-electron pulses and the sub-ensembles constituting the two-electron pulse. Specifically, this chirp decreases from a value around \SI{400}{fs/eV} at lower current to a value near or below \SI{300}{fs/eV} at higher current. In contrast, the two separate lobes of the two-electron distribution are displaced by an effective chirp of 130-140~fs/eV that only weakly depends on photocurrent (solid lines in Fig.~\ref{fig_4}b,d). First, the apparent independence of this inter-electron chirp on laser power illustrates that the strength of the Coulomb correlation is hardly influenced by stochastic interactions with electrons not entering the beam, as the correlation gap depends only weakly on the total photoelectron rate~\cite{Haindl2023a}. Second, the lower chirp value between the two lobes compared to that within the sub-ensembles reflects that the energy difference of the electrons emerges during propagation. This, in turn, causes reduced dispersion as compared to that experienced by variations in initial conditions upon photoemission. We note that these features of the experimentally retrieved longitudinal phase-space distributions of double-electron states are also reproduced by the trajectory simulations discussed in Fig.~\ref{fig_1}b. Considering further details of the distributions, we recognize a non-linear distortion in the experimentally reconstructed two-electron densities, flattening out towards zero delay and increasing towards larger absolute delays. This higher-order chirp (cf. Fig.~\ref{fig_4}b, and also Fig.~\ref{fig_3}f) suggests a correlation of the final electron energy difference with the remaining dispersion during propagation. Specifically, stronger scatterings tend to occur closer to the emitter, leaving a larger influence of dispersion after the interaction, while weaker interactions coincide with lower total dispersion.
\begin{figure*}[!t]
\centering
\includegraphics[width=\textwidth]{figures/phase_space_manuscript/fig_5/fig_5_v38.pdf}
\caption{Coherent shaping of two-electron pulses enables attosecond inter-particle correlations and coherent modulation of electron-light coupling. a,b) Experimental (a) and simulated (b) spectral pair histograms of energies $E_\mathrm{1/2} - E_0$ after IELS with a long laser pulse and experimental parameters $\Delta_\mathrm{L} \approx \SI{2}{ps}$,  $\hbar \omega_{\rm L}=\SI{2}{eV}$, $r_1 =\SI{1.9e-2}{}, r_2=\SI{1.6e-4}{}$ and   $|\beta|=0,0.3,0.6,0.9,1,1.2$.  c-f) Concept for coherent multi-particle excitation of quantum systems. c) Simulated temporal pair histogram at the sample plane. d) Upper panel: IELS-shaped two-electron Wigner function $W(t_1,t_2,E_1+\hbar \omega_{\rm L}/2,E_2+\hbar \omega_{\rm L}/2)$ (see Appendix \ref{aD}) at post-selection energy $(E_1-E_0,E_2-E_0)=(1.5,1.5)\hbar \omega_\mathrm{L}$ (circle) in the region indicated by the zoom-in. Lower panel: quasiprobability of the electron-electron temporal distance $W_{\omega_{\rm L}}(\Delta t,E_1,E_2)$ obtained from the time-convolution of the two-electron Wigner function (see scheme in the upper panel and Appendix \ref{aD}) for post-selection energies  $(1.5,-2.5)\hbar \omega_\mathrm{L}$ (hexagon), and $(-0.5,-3.5)\hbar \omega_\mathrm{L}$ (triangle). e,f) Interaction of two electrons with a second specimen with excitation frequency $\omega_{\rm L}$ placed after the IELS modulation and relaxing through the emission of a CL photon with equal energy (see sketch). The correlation signal (see Eq. (\ref{pclcoh})) is obtained after electron energy post-selection in temporal coincidence with the detection of a cathodoluminescence photon and is given by the multiplication of an incoherent contribution $2P_{\rm CL}J_{\omega_{\rm L}}$ (e) and a coherent modulation factor $\eta_{\omega_{\rm L}}$ (f).}
\label{fig_5}
\end{figure*}

To actively shape the two-electron time-energy correlations, we extend our scheme beyond the short-pulse regime. So far, we utilized IELS with laser pulses shorter than the single-electron ensemble duration for phase-space characterization. To realize a coherent phase modulation \cite{Feist2015} of the electron ensemble, we now employ IELS with a significantly longer laser pulse ($\Delta_{\rm L}\approx \SI{2}{ps}$), directly imprinting a fixed optical phase onto both electrons. Figure~\ref{fig_5}a shows the resulting phase modulation of electron pairs at photon energy $\hbar \omega_{\rm L}=\SI{2}{eV}$ and for various modulation strengths. We experimentally observe a checkerboard density pattern arising from a two-particle quantum walk, in agreement with simulations of classically (non-entangled) correlated electrons (see Fig.~\ref{fig_5}b and Appendix \ref{aC}). In contrast to the short-pulse IELS data (Fig.~\ref{fig_3}b\&c), here, the entire electron ensemble interacts with a fixed-phase light field over multiple optical periods, resulting in a multi-path interference of the measured sidebands~\cite{Feist2015} and a considerable population of highly scattered orders for both electrons, as shown in Fig.~\ref{fig_5}a.

The coherent interaction with the IELS field synchronously shapes the two-electron pulse such that each pair of final kinetic energies exhibits a different temporal correlation superimposed on the Coulomb-induced repulsion (Fig.~\ref{fig_5}c). This is exemplified by the two-electron Wigner function obtained after the IELS stage, shown in the upper panel of Fig.~\ref{fig_5}d for a fixed post-interaction energy coordinate $(E_1,E_2)$. Furthermore, by convoluting the two time coordinates of the pair Wigner function, we derive a quasiprobability (see Appendix \ref{aD}) that quantifies the electron current as a function of the time difference $\Delta t = t_1 - t_2$ and selected final electron energies, as indicated by the matching symbols in Fig.~\ref{fig_5}e\&f. Electrons post-selected on the gain side exhibit a quasiprobability peaked at $\Delta t$ values corresponding to integer multiples of the laser period, while postselection combinations of gain and loss yield maxima at intervals shifted by half an optical cycle. Both features result from the sub-cycle correlation among energy and arrival time optically induced on each electron. A dual-frequency arrival time correlation is consequently found when one of the two electrons is selected near an unscattered sideband.

These attosecond inter-particle correlations can be leveraged to coherently act on nanoscale systems in a manner controlled by the final post-selected energies. For example, the probability of detecting a free photon with energy $\hbar \omega_\mathrm{\rm CL}$, arising from the interaction of two-modulated electrons scattered at final energies $E_1$ and $E_2$ with a second sample showing coherent cathodoluminescence (CL) emission (e.g., transition or Cherenkov types of radiation \cite{paper149}), is given by (see Appendix \ref{aD})
\begin{align}
 P_{\rm \omega_{\rm CL}}(E_1,E_2) = 2P_{\rm CL}J_{\omega_{\rm CL}}(E_1,E_2)\, \eta_{ \omega_{\rm CL}}(E_1,E_2). \label{pclcoh}
\end{align}
The probability consists of an incoherent contribution $2P_{\rm CL}J_{\omega_{\rm CL}}$ and a coherent modulation factor $\eta_{ \omega_{\rm CL}}$. In Fig.~\ref{fig_5}e\&f, we compute both components for the two-electron states that are shown in the energy maps in Fig.~\ref{fig_5}a, with $\omega_{\rm CL}=\omega_{\rm L}$. The incoherent part is obtained by multiplying the fraction of events in which an uncorrelated electron pair emits a CL photon $2P_{\rm CL}$ by the number of pairs compatible with the outcome of the energy measurement, represented by the normalized two-electron energy density $J_{\omega_{\rm CL}}$ (Fig.~\ref{fig_5}e). On the other hand, the coherent component arises from the above-mentioned IELS-induced energy-dependent sub-cycle time correlations. These lead to a suppression or enhancement of CL emission via destructive or constructive interference of the total exciting field experienced by the sample (Fig.~\ref{fig_5}f) and relates to the Wigner function distributions in Fig.~\ref{fig_5}d. 
Without post-selection, the associated probability---obtained by integrating Eq. (\ref{pclcoh}) over the full energy domain---follows the superradiant scaling $2P_{\rm CL}(1+{\rm DOC}_{\omega_{\rm CL}})$, which depends on the so-called degree of coherence $0\leq {\rm DOC}_{\omega_{\rm CL}}\leq 1$~\cite{DiGiulio2021, Kfir2021a, GarciadeAbajo2021}. In particular, ensembles bunched through a combination of IELS modulation stages and free spatial drift yield a nonzero degree of coherence, whereas uncompressed ensembles show a vanishing ${\rm DOC}_{\omega_{\rm CL}}$~\cite{Yalunin2021,GarciadeAbajo2023}. In this regard, excitation control through post-selection relaxes spatial constraints between the IELS stage and the target system and enables a broader range of modulations, including emission suppression, in contrast to bunched beams, which only support signal amplification.


In summary, we mapped Coulomb-induced femtosecond-electronvolt splitting in multi-electron phase-space distributions and harnessed attosecond temporal correlations in an ultrafast transmission electron microscope.
To reconstruct these distributions, we extended the theoretical framework of quantum inelastic electron-light interaction to multi-electron states, incorporating classical ensembles averaging. 
In particular, we connected the spectrograms to the $N$-particle phase-space density via a simple multi-dimensional convolution (see Eq.~(\ref{conv1})). While the full distribution $\rho_N(\{t_i\},\{E_i\})$ is not accessible via single-pulse IELS, its marginals are obtained by filtering the gain part of the spectrum associated to each electron (see Eq.~(\ref{conv2})). 
We find that electrons emitted by the same laser pulse mutually exchange energy over centimeter-scale acceleration distances, thereby acquiring arrival time separations on the order of $\approx \SI{200}{fs}$ at the sample plane. Future studies employing $N$ femtosecond-separated laser pulses with adjustable delays will address the reconstruction of the full longitudinal $N$-electron phase-space distribution $\rho_N(\{t_i\},\{E_i\})$, provided that the energy gap exceeds the spectral resolution. By extending the duration of the IELS modulation, we imprint a coherent optical phase onto the entire multi-electron ensemble. We theoretically describe that the resulting multi-electron quantum walk can be employed to coherently control optical excitations in material systems without restrictive spatial requirements imposed by dispersive bunching.
More broadly, the framework of inelastic few-electron-light scattering opens various opportunities to modulate and control electron number states using light pulses tailored to the densities of the individual electrons and utilize density modulation to induce superradiance in bosonic or fermionic quantum systems.

\section*{Acknowledgements}

We acknowledge the continued support from the Göttingen UTEM team. We thank Thomas Rittmann for insightful discussions on phase-space density reconstruction. The experiments were conducted at the Göttingen UTEM Lab, funded by the Deutsche Forschungsgemeinschaft (German Research Foundation) through Project-ID 432680300 (SFB 1456, project C01 to C.R.), the Gottfried Wilhelm Leibniz program (RO 3936/4-1 to C.R.) and the European Union’s Horizon 2020 research and innovation programme under grant agreement no. 101017720 (FET-Proactive EBEAM). During the review stage of this manuscript, a preprint of related two-electron IELS experiments was posted \cite{Tzipermann2025}, and we acknowledge fruitful discussions with the authors.

\section*{Author contributions}

C.R. directed the study. R.H. conducted the experiments and data analysis with contributions from A.F. R.H. prepared the figures, with contributions from V.D.G. R.H. and V.D.G. conducted particle simulations. V.D.G. developed the theoretical framework. All authors discussed the results and interpretation. R.H., V.D.G., and C.R. wrote the manuscript with input and feedback from A.F.

\section*{Competing Interests}
The authors declare no competing interests.

\newpage

\section*{Appendix}

\appendix

\section{Trajectory simulations}
\label{aA}
The few-electron phase-space distributions were simulated with a relativistic particle trajectory simulation. In our model, we numerically integrate the equations of motion of particles mutually interacting via Coulomb's force and propagating in a static electric field
\begin{equation}
    \frac{\mathrm{d}\textbf{p}_i(t)}{\mathrm{d}t} = \frac{e^2}{4\pi\varepsilon_0}\,\sum_{j\neq i}^N\, \frac{\phantom{|}\rb_i-\rb_j\phantom{|^3}}{|\rb_i-\rb_j|^3} +e\Eb_\textrm{static}(\rb_i),
\end{equation}
where $\textbf{p}_i$, $\rb_i$, $e$ and $m_e$   are the relativistic momentum, position, charge, and rest mass of the $i\mathrm{-th}$ electron. Its velocity is calculated using the relation
$\textbf{p}_i = \gamma m_\mathrm{e} \textbf{v}_i = m \textbf{v}_i / \sqrt{1 - v_i^2/ c^2}$. The electrons are accelerated in a static two-stage electric field $E_\mathrm{static}(\rb_i)= E_\mathrm{ext}(\rb_i) + E_\mathrm{acc}(\rb_i)$. Here, the first summand corresponds to the non-linear extraction field 
$E_\mathrm{ext}(\rb) = U_\mathrm{ext} \rb / r^3 (r_\mathrm{tip}^{-1} - R_\mathrm{ext}^{-1})$
with extraction voltage $U_\mathrm{ext} = \SI{2}{kV}$, tip radius $r_\mathrm{tip}=\SI{1}{\mu m}$, and the distance of the extractor from the tip $R_\mathrm{ext} = \SI{500}{\mu m}$. The second summand is a constant axial accelerating electric field $E_\mathrm{acc}(\rb)=U_\mathrm{acc}/d_\mathrm{acc} \cdot \zz  \textrm{ if }  z \leq d_\mathrm{acc}$, with the acceleration voltage of the transmission electron microscope $U_\mathrm{acc} = \SI{198} {kV}$ and the accelerating distance $d_\mathrm{acc} = \SI{0.85}{m}$. For each simulation run, the electrons are randomly distributed on a 1$\mu$m-sized spherical tip within an area with radius \SI{170}{nm}. The finite pulse length of our photoemission laser and the finite initial spectral distribution of the electrons are incorporated in the simulation by a normal distributions of the start times and the initial velocities with standard deviations of $\sigma_{0,t}=\SI{80}{fs}$ and $\sigma_{0,E}=\SI{0.35}{eV}$, respectively. The ratio between the initial longitudinal $v_\parallel$ and transverse $v_\mathrm{trans}$ velocity components $v_\perp /  v_\parallel$ is given by $\tan(\theta)$, where $\theta$ is defined with respect to the normal of the tip surface at the place of emission and is uniformly distributed in the interval \SI{-2}{\degree} and \SI{+2}{\degree}. Finally, we account for a finite spectral resolution of our spectrometer and a finite stability of our high-voltage power supply by in incoherent temporal and spectral averaging with standard deviations $\sigma_{0,t}$=\SI{30}{fs} and  $\sigma_{0,E}$=\SI{0.13}{eV}, respectively.

\section{Multi-electron IELS theory for short laser pulses}
\label{aB}
The theoretical description of the energy correlation maps showed in the main text can be carried out through a $N$-electron generalization of the well-established formulation of IELS experiments performed in electron microscopes with acceleration voltages in the kV range \cite{PLZ10,paper151,paper311}. In this regime, the minimal coupling Hamiltonian, describing the interaction of $N$ swift electrons traveling with velocities $\{\vb_i\}$ with a classical time-varying vector potential $\Ab(\rb,t)$, can be simplified by linearizing the electrons' dispersion relation around a single central momentum $\kb_0$, taken to be the average momentum, corresponding to the relativistic energy $E_{0}=c\sqrt{\me^2 c^2 + \hbar^2 k_{0}^2}$ and to the velocity $\vb=(\hbar c^2/E_{0})\kb_{0}$. By further neglecting spin flips, ponderomotive forces, small differences of the order $|\vb-\vb_i|/c\sim 10^{-5}$ between electrons' velocities, and Coulomb repulsion close to the Si membrane, we can solve the Schr\"{o}dinger equation $\ii \hbar \partial_t \psi^i(\rb_i ,t)=\mathcal{H}^i\psi^i(\rb_i,t)$ for each $i$-electron separately with Hamiltonian $\mathcal{H}^i=E_{0}-\hbar \vb \cdot(\ii \nabla_{\rb_i}+\kb_{0})+(e\vb/c)\cdot \Ab(\rb_i,t)$, which admits the analytical solution  
\begin{align}
 \psi^i(\rb_i,t)&=\psi^i_0(\rb_i,t)\label{pinem1}\\
 &\times \exp\Bigg\{(-\ii e/\hbar c) \int_{-\infty}^{t} dt' \, \vb \cdot \Ab[\rb_i-\vb(t-t'),t'] \Bigg\} \nonumber 
\end{align}
for an initial wave function $\psi^i_0(\rb_i,t)=\ee^{ \ii(\kb_0\cdot \rb_i - E_{0} t/\hbar)} \phi^i_0(\rb_i-\vb t)$ entering the interaction zone. We remark that by separately treating the problem for each electron, we tacitly exclude the presence of quantum correlations, i.e., entanglement, in the beam. We support this assumption by arguing that stochastic interaction with electrons that are not coupled into the column would likely wash out any Coulomb-induced coherent correlation, whose preservation would require a design of the photoemission and propagation process beyond the current state-of-the-art setup. Furthermore, an analysis similar to the one in Ref.~\cite{Tzipermann2025} of the data shown in Fig.~\ref{fig_5}a yielded a lack of longitudinal energy entanglement of two-electron states at the sample plane. By now focusing our attention on a semi-monochromatic laser pulse at frequency $\omega_{\rm L}$ for which the vector potential can be described by a slowly varying electric-field envelope with a negligible time derivative $\vec{\mathcal{E}}(\rb,t)$ as $\Ab(\rb,t)=(-\ii c/\omega_{\rm L})\vec{\mathcal{E}}(\rb,t)\ee^{-\ii \omega_{\rm L} t}+\,{\rm c.c.}$. Eq. (\ref{pinem1}) can be reformulated in terms of a superposition of amplitudes describing the loss and gain of $\ell$ light quanta ($\ell>0$ for gain and $\ell<0$ for loss) by each individual electron via the Jacobi-Anger expansion $\ee^{\ii u \sin \theta}=\sum_\ell J_\ell (u)\ee^{\ii \ell \theta}$ [Eq.
(9.1.41) of Ref. \citenum{AS1972}]. By doing so, Eq. (\ref{pinem1}) transforms into
\begin{align}
 \psi^i(\rb_i,t)=\psi^i_0(\rb_i,t)\!\!\sum_{\ell=-\infty}^\infty \!J_{\ell}(2|\beta(\rb,t)|)\,\ee^{\ii \ell \varphi(\rb_i,t)},\label{pinem2}
\end{align}
with $\varphi(\rb_i,t)={\rm arg}\{-\beta(\rb_i,t)\}$ written in terms of the time-dependent coupling coefficient $\beta(\rb,t)=(e/\hbar \omega_{\rm L})\int_{-\infty}^t dt' \,\vb\cdot \vec{\mathcal{E}}[\rb-\vb(t-t'),t]\,\ee^{-\ii\omega_{\rm L} t'}$. We now take the electrons to propagate along the $z$-axis while modeling the laser pulse, arriving at time $t_{\rm L}$ at the sample, with the temporal Gaussian profile $\vec{\mathcal{E}}(\rb,t)=\vec{\mathcal{E}}(\rb)\ee^{-(t-t_{\rm L})^2/2\Delta_{\rm L}^2}$. For the duration of the laser pulses (tens of fs), electron coherence times (few fs), and interaction times (sub-fs), used in this work, we can approximate $\beta(\rb,t)\approx \ee^{-[z/v-(t-t_{\rm L})]^2/2\Delta_{\rm L}^2-\ii \omega_{\rm L} (t-z/v)}\beta$ with $\beta=(e/\hbar \omega_{\rm L})\int_{-\infty}^z dz'\, \mathcal{E}_z(\Rb,z')\,\ee^{-\ii \omega_{\rm L} z'/v}$. By plugging $\beta(\rb,t)$ into Eq. (\ref{pinem2}), we obtain
\begin{align}
\psi^i(\rb_i,t)=\psi_0^{i}(\rb_i,t)\mathcal{P}(\Rb_i,z_i-vt)\nonumber 
\end{align}
where the IELS amplitude $\mathcal{P}(\Rb,z)=\sum_{\ell=-\infty}^\infty J_{\ell}\big[2\,\ee^{-(z/v-t_{\rm L})^2/2\Delta_{\rm L}^2}|\beta|\big]\ee^{\ii \ell (\omega_{\rm L} z/v+\varphi)}$ and $\varphi={\rm arg}\{-\beta\}$. The probability $P(\{E_i\})=\prod_i \Gamma_{i}(E_i)$ of measuring the $N$ electrons in a set of energies $\{E_i\}$ is directly connected to their wave functions via the one-dimensional spatial Fourier transform $\Gamma_{i}(E_i)=(1/2\pi \hbar v)\int d\Rb \big|\int_{-\infty}^\infty dz\,\psi^{i}(\rb) \ee^{-\ii [k_0+(E_i-E_0)/v\hbar ]z}\big|^2$.

For a Gaussian electron of main energy $E_{0,i}$ and coherence time $\Delta_c$, focused around the transversal coordinate $\Rb_0$, and crossing the $z_{0,i}=t_{0,i}/v$ plane at $t=0$, we can write $\psi^{i}_{0}(\rb)=f(\Rb-\Rb_0)\,\ee^{\ii k_{0,i}z-(z-z_{0,i})^2/4v^2\Delta_{\rm c}^2}/(2\pi v^2\Delta_c^2)^{1/4}$ with $f(\Rb-\Rb_0)$ a function satisfying $|f(\Rb-\Rb_0)|^2=\delta(\Rb-\Rb_0)$. For such state, the Fourier transform can be carried out exactly by using the expansion $J_\ell(x)={\rm sign}^\ell \!\ell\sum_{m=0}^\infty (-1)^m (x/2)^{2m+|\ell|}/m!(m+|\ell|)!$. The resulting expression can be further simplified by neglecting terms of order $(\Delta_{\rm c}/\Delta_{\rm L})^4$ and considering $\omega_{\rm L}\Delta_c\gg 1$, and is given by $\Gamma_i(E_i)=\Gamma(t_{\rm L}-t_{0,i},E_i-E_{0,i})$ with
\begin{align}
\!\!\Gamma(t,E) =\frac{\Delta_c}{\hbar}\!\!\!\!\sum_{\substack{\ell=-\infty\\m,m'=0} }^\infty \!\!\!\!b_{mm' \ell}\, \ee^{- n t^2/\lambda \Delta_{\rm L}^2-2[E/\hbar-\ell\omega_{\rm L}]^2\Delta_{\rm c}^2},\label{gi}
\end{align}
where $\lambda =1+4 n\Delta_{\rm c}^2/\Delta_{\rm L}^2 $, $n=|\ell|+m+m'$, and $b_{mm'\ell}=(-1)^{m+m'}|\beta|^{2n}\sqrt{2}/ \sqrt{\pi\lambda} m!m'! (|\ell|+m)!(|\ell|+m')!$.

Finally, the recorded energy maps $\langle \Gamma(t_{\rm L},\{E_i\})\rangle$ will be the result of the incoherent average of the $N$-electron IELS spectra $P(\{E_{i}\})=\prod_{i=1}^N \Gamma(t_{\rm L}-t_{0,i},E_i-E_{0,i})$, acting as convolution function, over the classical $\rho_{N}(\{t_{0,i}\},\{E_{0,i}\})$ ensemble of delays $\{t_{0,i}\}$ and energies $\{E_{0,i}\}$ at the membrane plane, namely, Eq. (\ref{conv1}). The factor $1/N$  originates from the normalization condition $N=\int \{dE_i\}\{dt_i\} \rho_N(\{t_i\},\{E_i\})$. In particular, the simulated maps in Fig. \ref{fig_3}b\&c have been computed by using the two-electron phase-space density obtained from the trajectory simulations [see Fig. \ref{fig_1}] while setting $\Delta_{\rm L}=72$ fs, $\Delta_{\rm c}=1.29$ fs, and $|\beta|=5.68$. The coherence time is derived from the spectral width of the non-stochastically broadened single-electron spectrum of $\hbar \sqrt{2\log{2}}/\Delta_{\rm c}\approx 0.6$ eV acquired at very low photoexcitation laser powers \cite{Feist2017}.
 
\section{Slicing and reconstruction algorithm}
\label{aC}
Because indistinguishable, the two-electron phase-space density $\rho_2(t_{1},t_{2},E_{1},E_{2})$ will be symmetric under exchange of electron labels and therefore its marginalized distribution can be expressed as the sum of two contributions $\rho_2(t,E)=\rho_A(t,E)+\rho_B(t,E)$. In particular, because of Coulomb repulsion, these components will be localized in energy and time around the values $(\bar{E}_{A},\bar{t}_{A})<(\bar{E}_{B},\bar{t}_{B})$ [see Fig. \ref{fig_1}]. The slicing procedure leverages such time and energy localization to extract them from different parts of the spectrograms. Indeed, under the assumption that $\Delta_{\rm L}$ is small compared to the temporal separation between $\rho_A$ and $\rho_B$, one can integrate over $E_1$ the maps $\langle \Gamma(t_{\rm L},E_1,E_2)\rangle $ for each value of $E_2$ in the green (for $t_{\rm L}\sim \bar{t}_{B}$) and pink (for $t_{\rm L}\sim \bar{t}_{A}$) top selection regions of Fig. \ref{fig_3}b\&c (or equivalently swapping the roles of $E_1$ and $E_2$ for the rightmost selection regions), to reduce Eq. (\ref{conv1}) into the two one-dimensional convolutions reported in Eq. (\ref{conv2}). 

To obtain the marginalized phase-space densities, we make two other approximations: {\it (i)} short laser-pulse duration compared to the temporal extension of the distributions ($\Delta_{\rm L}/\sqrt{n}\sim 30$ fs for the outermost energy sidebands); {\it (ii)} the energy and time dependencies can be expressed in the form  $\rho_{A/B}(t,E)=\eta_{A/B}(t)\,\ee^{-[E-E^{\rm chirp}_{A/B}(t)-\bar{E}_{A/B}]^2/2\sigma^2_{\rm inc}}/\sqrt{2\pi \sigma^2_{\rm inc}}$, i.e., an incoherent Gaussian energy envelope of standard deviation $\sigma_{\rm inc}$ with chirp function $E^{\rm chirp}_{A/B}(t)$ multiplied by an unknown time-dependent distribution $\eta_{A/B}(t)$ is assumed. After plugging this form into Eq. (\ref{conv2}) and by using Eq. (\ref{gi}) in the energy regions of interest, we obtain
\begin{align}
    I_{A/B}(t_{\rm L},E_2)&\approx   C\,\eta_{A/B}(t_{\rm L}) \label{I2}\\
   & \times\sum_{\ell=\ell_{\rm min}}^{\ell_{\rm max}}\!\! a_\ell \, \ee^{-[E_2-E_{A/B}^{\rm chirp}(t_{\rm L})-\bar{E}_{A/B}-\ell \hbar \omega_{\rm L}]^2/2\sigma^2}, \nonumber
\end{align}
where $\sigma^2=\hbar^2/4\Delta_{\rm c}^2+\sigma_{\rm inc}^2$, $C$ and $a_\ell$ are real-valued coefficients, and the summation over $\ell$ has been limited to the range $[\ell_{\rm min},\ell_{\rm max}]$ to account for the finite number of selected sidebands. 
Finally, $E_{A/B}^{\rm chirp}(t)$ is extracted from the phase of the Fourier transform of Eq. (\ref{I2}), performed along the energy axis and subtracting a constant phase such that $E_{A/B}^{\rm chirp}(\bar{t}_{A/B})=0$, while, $\sigma_{\rm inc}$ is obtained by fitting the envelope of the absolute value of the Fourier transform to a Gaussian function. The temporal profile $\eta_{A/B}(t)$ follows after the integration of Eq. (\ref{I2}) over $E_2$ and by normalizing the result. 

\section{Multi-electron coherent modulation of material excitations}
\label{aD}
To display the energy and time correlations formed during an IELS interaction, we compute the two-electron Wigner function (see Fig.~\ref{fig_5}d)
\begin{align}
W(t_1,t_2,E_1,E_2)=\langle W_1(t_1,E_1)W_2(t_2,E_2)\rangle,\label{twowig}
\end{align}
obtained by averaging (indicated by the $\langle \cdot \rangle$ symbol) the multiplication of two single-electron Wigner functions $W_i(t_i,E_i)=(1/2\pi)\int_{-\infty}^\infty dy\, \psi^i(vt_i+y/2)\psi^i(vt_i-y/2)\ee^{\ii [k_0+(E_i-E_0)/\hbar v]y}$ \cite{W1932} over the ensemble of delay times and initial energies $\rho_2(t_{0,1},t_{0,2},E_{0,1},E_{0,2})$. For electrons exiting the IELS interaction region and for $\Delta_{\rm L}/\sigma\gg 1$, we have
\begin{align}
\!\!W_i(&t_i,E_i)=\frac{1}{\pi}\ee^{-(t_i-t_{0,i})^2/2\Delta_{\rm c}^2}\!\!\! \!\!\sum_{\ell,\ell'=-\infty}^\infty \! \!\! \!\!\!J_\ell(2|\beta|)J_{\ell'}(2|\beta|)\label{wie}\\
&\times \ee^{-[(E_i-E_{0,i})/\hbar \omega_{\rm L}-(\ell+\ell')/2]^2 2\omega_{\rm L}^2 \Delta_{\rm c}^2}\ee^{\ii (\ell-\ell')\omega_{\rm L} (t_i+\varphi/\omega_{\rm L})},\nonumber
\end{align}
which we use as the basis for the calculation of Eq. (\ref{twowig}), shown in Fig.~\ref{fig_5}d. Ensemble averages, such as the one appearing in Eq. (\ref{wie}), have been computed by taking the arithmetic mean of the quantities in brackets, evaluated at the time delays ${t_{0,i}}$ and initial energies ${E_{0,i}}$ obtained from the Monte Carlo simulations (see Appendix \ref{aA}). In this regard, our simulations provide the necessary sampling of the relevant phase space, defined by $\rho_2(t_1,t_2,E_1,E_2)$, to accurately approximate the averaging integral in Eq. (\ref{wie}).

We theoretically derive the probability in Eq. (\ref{pclcoh}) of exciting a second sample that, relaxing back to the ground state, emits a CL photon measured at frequency $\omega_{\rm CL}$ in coincidence with the detection of an energy pair $(E_1,E_2)$ (see sketch in Fig. \ref{fig_5}). Its evaluation requires the knowledge of the density of electrons compatible with the CL emission and the energy measurement 
\begin{align}
J_{\omega_{\rm CL}}(E_1,E_2)=\frac{1}{2}\langle \PM^1_0 &(E_1+\hbar \omega_{\rm CL})\PM^2_0(E_2)\label{J}\\
&+\PM^1_0(E_1)\PM^2_0(E_2+\hbar \omega_{\rm CL})\rangle \nonumber
\end{align}
as well as of the modulation factor
\begin{align}
\eta_{\omega_{\rm CL}}(E_1,&E_2)=1+{\rm Re}\{\langle \PM^1_{\hbar \omega_{\rm CL}}\!(E_1+\hbar \omega_{\rm CL}/2)\label{eta}\\
&\times \PM^{2*}_{\hbar \omega_{\rm CL}}\!(E_2+\hbar \omega_{\rm CL}/2)\rangle \}/J_{\omega_{\rm CL}}(E_1,E_2)\nonumber,
\end{align}
written in terms of the projected coherence factor (PCF) $\PM^i_{E}(E')=\int_{-\infty}^\infty dt\, W_i(t,E')\,\ee^{\ii E t/\hbar}/\hbar$ (see Ref.~\citenum{DHR25}). We notice that the multiplication of two Fourier transforms appearing in Eq.~(\ref{eta}) can also be evaluated through the temporal Fourier transform of the convolution $W_{\omega_{\rm CL}}(\Delta t,E_1,E_2)=\int_{-\infty}^\infty dt_2 W(\Delta t + t_2,t_2,E_1+\hbar \omega_{\rm CL}/2,E_2+\hbar \omega_{\rm CL}/2)$ (see bottom panel in Fig.~\ref{fig_5}d).
\clearpage

\clearpage

\twocolumngrid

\bibliography{refs, pinem_correlation_manuscript}
\bibliographystyle{apsrev4-2}
\end{document}